# Oxidized-monolayer Tunneling Barrier for Strong Fermi-level Depinning in Layered InSe Transistors

Yi-Hsun Chen[1], Chih-Yi Cheng[1], Shao-Yu Chen[1,2], Jan Sebastian Dominic Rodriguez[3], Han-Ting Liao[4], Kenji Watanabe[5], Takashi Taniguchi[5], Chun-Wei Chen[4], Raman Sankar[6,7], Fang-Cheng Chou[6], Hsiang-Chih Chiu[3] and Wei-Hua Wang[1*]

[1]Institute of Atomic and Molecular Sciences, Academia Sinica, Taipei 106, Taiwan

[2]School of Physics and Astronomy, Monash University, Victoria 3800, Australia

[3]Department of Physics, National Normal Taiwan University, Taipei 106, Taiwan

[4]Department of Materials Science and Engineering, National Taiwan University, Taipei 106, Taiwan

[5]Advanced Materials Laboratory, National Institute for Materials Science, Tsukuba, 305-0044, Japan

[6]Center of Condensed Matter Sciences, National Taiwan University, Taipei 106, Taiwan

[7]Institute of Physics, Academia Sinica, Taipei 115, Taiwan

ABSTRACT

In 2D-semiconductor-based field-effect transistors and optoelectronic devices, metal–semiconductor junctions are one of the crucial factors determining device performance. The Fermi-level (FL) pinning effect, which commonly caused by interfacial gap states, severely limits the tunability of junction characteristics, including barrier height and contact resistance. A tunneling contact scheme has been suggested to address the FL pinning issue in metal–2D-semiconductor junctions, whereas the experimental realization is still elusive. Here, we show that an oxidized-monolayer-enabled tunneling barrier can realize a pronounced FL depinning in indium selenide (InSe) transistors, exhibiting a large pinning factor of 0.5 and a highly modulated Schottky barrier height. The FL depinning can be attributed to the suppression of metal- and disorder-induced gap states as a result of the high-quality tunneling contacts. Structural characterizations indicate uniform and atomically thin surface oxidation layer inherent from nature of van der Waals materials and atomically sharp oxide–2D-semiconductor interfaces. Moreover, by effectively lowering the Schottky barrier height, we achieve an electron mobility of 2160 cm$^2$/Vs and a contact barrier of 65 meV in two-terminal InSe transistors. The realization of strong FL depinning in high-mobility InSe transistors with the oxidized monolayer presents a viable strategy to exploit layered semiconductors in contact engineering for advanced electronics and optoelectronics.

INTRODUCTION

Two-dimensional (2D) semiconductors present great potential for electronics and optoelectronics applications due to several unique characteristics, including efficient electrostatics, a lack of short-channel effects, and the absence of dangling bonds.[1-6] Recent studies revealed that one of the challenges of fabricating 2D-semiconductor-based transistors is to achieve an Ohmic contact at the interface of the electrodes and the 2D semiconductors.[7-10] In conventional bulk semiconducting devices, the Ohmic contact can be realized by matching the work function of the metals to the bands of the semiconductors.[11-13] However, Changsik Kim et al.[14] have shown a strong Fermi Level (FL) pinning in 2D semiconductor transistors; i.e. the Schottky barrier height (SBH), $\Phi_{SB}$, is virtually independent of the work function of contact metals, resulting in a lack of tunability of contact resistance as well as the low field effect mobility and output current.[6, 15] Moreover, the polarity of carrier transport cannot be manipulated by varying contact metals,[6] critically limiting the controllability of carrier extraction in 2D semiconductor transistors. Mitigating the FL pinning effect is thus an urgent need to manifest the intrinsic nature of 2D semiconductors, to modulate the carrier transport properties, and to improve transistor performance.

A tunneling contact has been suggested to address the FL pinning effect in 2D semiconductor devices by reducing: (i) disorder-induced gap states (DIGS),[6, 16] (ii) metal-induced gap states (MIGS),[17-19] and (iii) interface dipoles.[20] By inserting an insulating layer between the metal and the semiconductor, the MIGS and interface dipoles can be effectively reduced by blocking the metallic wave function[17-19] and by neutralizing interface dipoles,[20] respectively. In the process of the device fabrication, the insulating layer can protect the underlying 2D semiconductors from direct bombardment and metal diffusion during the metal deposition, decreasing the DIGS in the metal–2D semiconductor interface.[6, 16] Previous studies have used metal/insulator/semiconductor (MIS) structure to tune the $\Phi_{SB}$ in 2D semiconductor devices.[15, 20, 21] However, these studies presented limited FL depinning, suggesting the difficulties to achieve a high-quality tunneling layer at the metal–semiconductor junctions. We overcome these issues by utilizing a surface oxidation layer (OL) in 2D semiconductors as an effective tunneling barrier. In the in-plane direction, a natively grown OL can yield a high compatibility between the oxide and layered semiconductor,[22, 23] enabling a large-scale and atomically smooth oxide functioned as a tunneling contact in 2D semiconductor devices. In the out-of-plane direction, because of the nature of van der Waals crystals, the thickness of the oxide layer can be precisely controlled down to atomic-scale resolution, facilitating a high tunability of tunneling efficiency.

In this work, we address the FL pinning effect in 2D-semiconductor-based transistors by demonstrating high-performance indium selenide (InSe) field-effect transistors with an atomically thin, precisely controlled tunneling barrier made by a surface OL. InSe is an emerging 2D semiconducting material with potential applications because of its extraordinary intrinsic characteristics, including a small effective mass in the conduction band, small optical bandgap (1.25 eV), high Hall mobility,[24, 25] and weak electron-phonon scattering.[26] Hence, one envisions in InSe a channel material with high electronic performance that validates the demonstration of FL depinning. Here, we fabricate the atomically thin OL by a UV-ozone process that is highly scalable and more compatible with semiconductor manufacturing processes. By employing the surface OL as part of the tunneling contact, we successfully achieve a strong FL depinning in InSe transistors and realize a high pinning factor of 0.5. Moreover, the barrier height and the threshold voltage can be effectively modulated by varying the work function (WF) of the contact metal because of the pronounced FL depinning. Accordingly, with the selected WF of the contact metal, we achieve a low contact barrier of 65 meV in the InSe devices, enabling a high two-terminal electron mobility of 2160 cm$^2$/Vs.

RESULTS AND DISCUSSION

The InSe samples are mechanically exfoliated from InSe crystals onto a highly-doped Si substrate with a 300 nm SiO$_2$ dielectric layer under ambient conditions. After exfoliation, we grown the surface OL on InSe flakes by employing UV ozone treatment at 80 °C for 10 seconds. The atomic structure and chemical properties of the surface OL are examined by cross-sectional transmission electron microscopy (TEM) and energy-dispersive X-ray spectroscopy (EDX) mapping, as shown in Figure 1a. Remarkably, the OL is uniformly grown, and the underlying InSe layers remain of a high-quality crystallinity, highlighting an atomically sharp and clean interface. The surface OL is amorphous with a thickness of 1 nm, approximately 1.3 times that of the single-layer InSe, implying that only the topmost monolayer is oxidized. The EDX mapping, which has been used to estimate the relative abundance of elements in the heterostructures, reveals that a substantial amount of Se atoms are substituted by O atoms within the OL regime, which is further supported by X-ray photoelectron spectroscopy data (Supplementary Fig. S1). The In concentration remains constant up to the surface, suggesting the formation of substoichiometric InO$_x$. The TEM and EDX mapping clearly indicate the formation of a uniform OL generated from the topmost InSe monolayer, which we denote as an oxidized monolayer (OML) from here forward. We note that in contrast to natively grown oxide (Supplementary Fig. S3) and oxygen plasma,[27] UV ozone facilitates the strong oxidation of the InSe surface due to the presence of singlet oxygen atoms, which are chemically active.[28] The strong oxidizability could in principle lead to the domination of the vertical oxidation reaction over the lateral reaction triggered by the defects,[29] resulting in a uniform OML observed in our InSe samples.

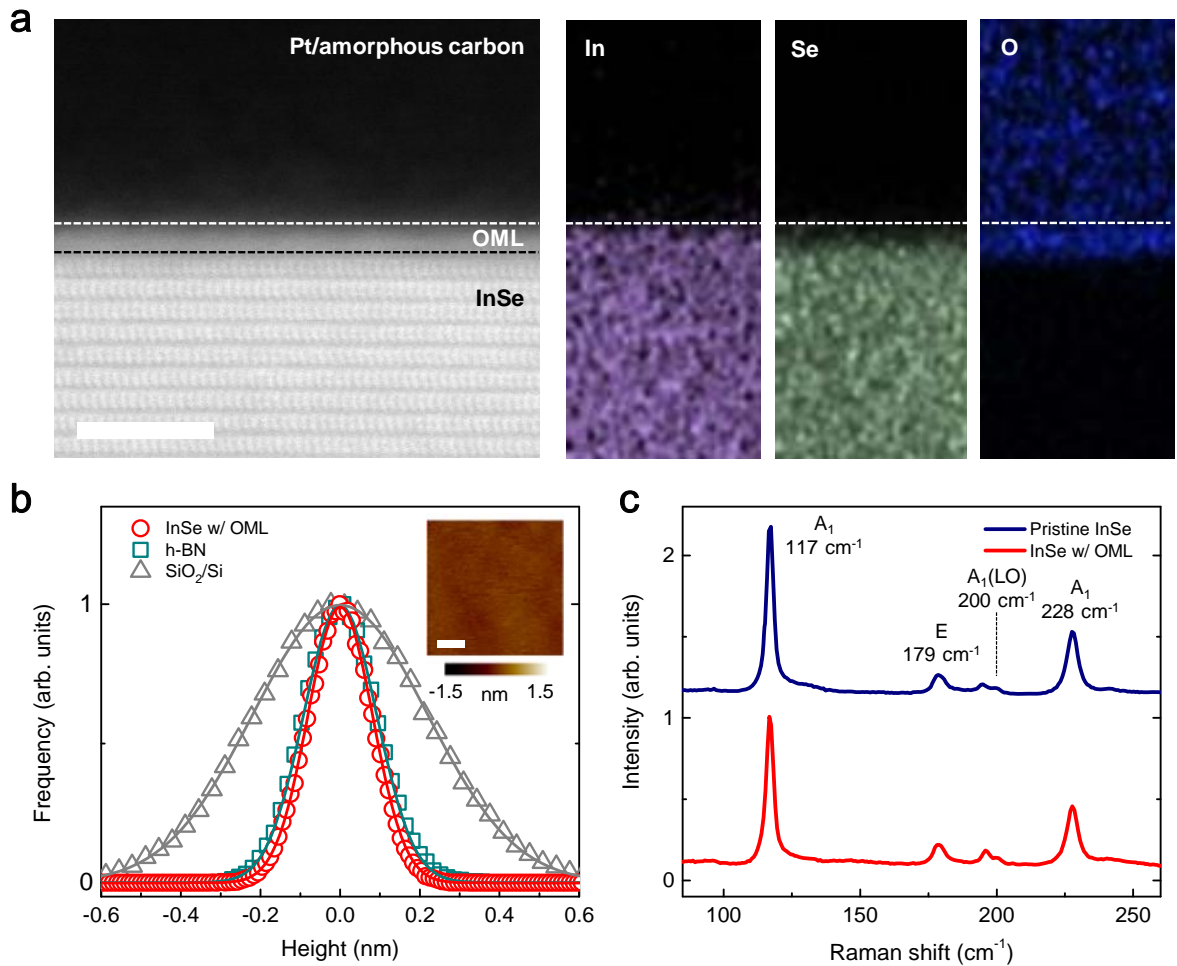

**Figure 1**. (a) Cross-sectional transmission electron microscopy and energy-dispersive X-ray spectroscopy mapping images of an InSe sample processed by UV ozone, showing an atomically sharp, clean interface and high-crystalline-quality InSe layers underneath the surface OML. The scale bar is 5 nm. (b) A histogram of the height distributions for the surface OML of an InSe

sample (red), an h-BN bulk sample (blue), and the SiO$_2$/Si substrate (green). The solid lines are curves fitted by a Gaussian distribution. The surface roughness values of the InSe surface OML, h-BN bulk, and SiO$_2$/Si substrate are 81 pm, 91 pm, and 219 pm, respectively. Inset: an AFM image of the surface OML of an InSe sample, showing an ultrasmooth surface morphology. The scale bar is 100 nm. (c) A comparison of the Raman spectra of pristine InSe (blue curve) and the InSe sample with the OML (red curve).

The surface topography of the OML of InSe samples is characterized by using atomic force microscopy (AFM). The AFM images were acquired by the Peak-Force tapping mode using a sharp probe with a tip radius of 2 nm. (See Supplementary Information S2 for detail information.) The inset of Figure 1b shows the AFM image of the OML that manifests a very uniform surface, signifying that even with a thickness of 1 nm, the surface OL is highly compatible with the underlying InSe. Figure 1b compares the histogram of the OML surface with that of a bulk hexagonal boron nitride (h-BN) flake and a SiO$_2$/Si substrate. Notably, the OML surface exhibits a very small roughness of 80 pm, confirming the high compatibility of the OML with the underlying InSe. It is noteworthy that the roughness of the oxidized InSe surface is comparable to that of h-BN (i.e., 90 pm), which exhibits an atomically smooth surface[30] and can be used in tunneling contacts for 2D-semiconductor-based devices.[31, 32] This ultra-uniform surface thus suggests the viability of employing the OML as a tunneling barrier. As a comparison, nucleation sites are observable in the InSe samples not subjected to the UV-ozone treatment, which may be attributed to the native oxide that arises during exposure to the ambient atmosphere (Supplementary Fig. S3 and S4). Therefore, the uniform surface of the OML indicates a passivation and a stabilization effect resulting from the chemically stable surface oxide.[33]

We further perform Raman spectroscopy to understand the crystal quality of the InSe sample before and after the UV-ozone treatment. As shown in Figure 1c, we observe four characteristic Raman peaks in the pristine InSe sample at 117, 179, 200, and 228 cm$^{-1}$, corresponding to the out-of-plane ($A_1$) and in-plane ($E$) vibration modes.[34] After the UV-ozone treatment, these four characteristic Raman peaks show insignificant changes to their peak intensity and energy. The full width at half maxima before (after) the UV-ozone treatment are 2.9 (3.1), 5.7 (5.1), 2.3 (2.7), and 4.7 (4.8) for Raman modes at 117, 179, 200, and 228 cm$^{-1}$, respectively. The invariance of these Raman modes before and after the oxidation process clearly confirms that the crystal quality of the underlying InSe is negligibly affected by the UV-ozone treatment. No additional peaks corresponding to the oxide are observed, suggesting that the oxide is amorphous. To summarize, the comprehensive characterizations of TEM, AFM, and Raman spectroscopy show strong evidence of the ultra-uniform OML interfaced with the high-quality layered InSe underneath.

To understand the effects of introducing the ultra-smooth and uniform OML on the electrical properties of the InSe transistors, we fabricate back-gated devices with the contact structure illustrated in Figure 2a. The thickness of the InSe flakes used for the devices ranges from 10 to 20 nm, as confirmed by AFM. The InSe samples are then transferred onto h-BN flakes by a dry transfer technique. We employ the h-BN flakes as the substrate for the InSe channel to reduce the extrinsic carrier scattering sources at the bottom interface (Supplementary Fig. S5).[35] The InSe/h-BN stacking structures are annealed in a furnace at 300 °C with mixed gas (95% Ar and 5% H$_2$) to improve the quality of the InSe/h-BN interface and to remove the residual polymer at the surfaces. These InSe/h-BN samples are then oxidized by the UV-ozone process at a temperature ranging from 80 to 100 °C for 10 seconds, followed by depositing electrical contacts (contact metal/Au of 15 nm/40 nm) with e-beam evaporation at a base pressure of $1 \times 10^{-7}$ Torr. To attain high uniformity of the tunneling barrier, we employ a resist-free method to define the source and drain contacts[36] to avoid the potential resist residue in conventional lithography techniques, allowing us to study the FL depinning effect accurately. For the resist-free fabrication method, we use the commercially available TEM grids to define the device channel by the shadow effect with the width ranging from 5 to 12 μm. The TEM grids are attached to a premade holder, which are then mounted on a home-made aligner. We then align the InSe flakes to the bar of the TEM grid under an optical microscope, subsequently followed by the metal deposition. We note that the TEM grid is not in contact with the InSe flake during the whole process, leading to the pristine condition of the InSe surface. The inset of Figure 2b shows an optical microscopy (OM) image of a typical InSe device. All the electrical characterization are performed in a Physical Property Measurement System in a helium environment at a pressure of 20 Torr.

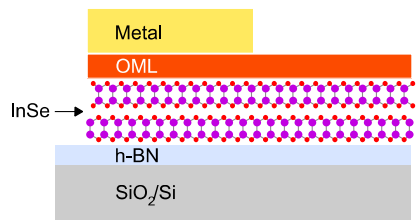

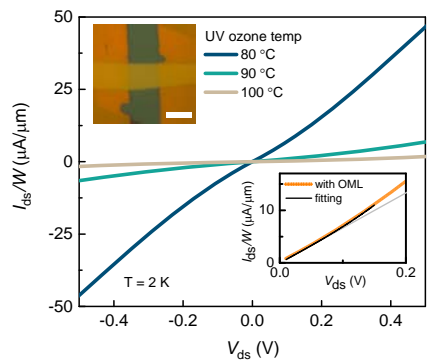

**Figure 2.** (a) A schematic of the InSe device structure at the contact regime shows a surface oxidation layer embedded in between the contact metal and InSe. (b) The output curves of InSe devices utilizing UV-ozone processing temperatures of 80, 90 and 100 °C for a duration of 10 seconds. The $V_{gs} = 80$ V, and the measured $T = 2$ K. Inset: (Upper left) An OM image of a typical InSe device. The scale bar is 10 μm. (Lower right) The output curve for the InSe device utilizing a UV-ozone temperature of 80 °C for 10 seconds, and the fitting curve calculated by Simmons' model. The gray line is a linear extrapolation of the output curve from the low-voltage regime.

We first present the tunneling transport behavior of the OML-embedded InSe devices. As shown in Figure 2b, the InSe devices constructed with UV-ozone processing temperatures of 80, 90, and 100 °C exhibit nonlinear output curves ($I_{ds} - V_{ds}$ curves). These nonlinear output curves can be well described by Simmons' model that manifests a cubic term in the $V_{ds}$ dependence (inset in Figure 2b and Supplementary Fig. S6), suggesting that the tunneling effect governs the charge transport as a result of introducing the OML. For higher processing temperatures, the $I_{ds}$ decreases because of a larger contact resistance corresponding to a thicker OL, which is consistent with the tunneling transport behavior in the OML-embedded InSe devices. The large $I_{ds}$ of the InSe device with an OML grown at 80 °C indicates the optimal tunneling efficiency with a low contact barrier, therefore, this condition is employed to fabricate the OML for subsequent studies. Moreover, compared with the InSe sample without the OML, the InSe devices embedded with the OML exhibit a higher $I_{ds}$ and lower barrier height (Supplementary Fig. S7), which is consistent with the tunneling contact behavior.[20]

To evaluate the SBH of the MIS contact of the InSe devices, we conduct temperature ($T$)-dependent transport measurements and extract the SBH under flat-band gate bias conditions. Based on the thermionic emission model,[9] the drain current injected through a reverse-biased Schottky barrier can be written as:

$$I_d = A^*T^{3/2} \exp\left[-\frac{E_A}{k_B T}\right] [1 - \exp(-\frac{qV_{ds}}{k_B T})], \quad (1)$$

where $A^*$ is Richardson's constant, $k_B$ is the Boltzmann constant, $T$ is the temperature, $q$ is the elementary charge, $E_A$ is the activation energy, and $V_{ds}$ is the source–drain voltage. By fitting the Arrhenius plot of $I_d/T^{3/2}$, we can extract $E_A$, which represents the barrier height that the carriers have to overcome. The $V_{gs}$ dependence of $E_A$ is plotted to determine the SBH of the devices under the flat-band condition (Supplementary Fig. S8). The SBH extraction is found to only be applicable in the temperature range from 300 to 400 K, where the thermionic emission current dominates the carrier transport.[14] Moreover, we note that the extracted SBH corresponds to an effective SBH that represents the overall contact behavior because the insulating layer is not considered in the current–voltage characteristics described in equation (1).

We now discuss how the OML-embedded MIS contact affects the FL pinning effect in the InSe devices by examining the pinning factor and charge neutrality level.[37] The relation between the SBH and the metal WF can be depicted by the following equation:

$$\phi_{Bn} = S(\phi_m - \phi_{CNL}) + (\phi_{CNL} - \chi) = S\phi_m + b \quad (2)$$

where $\phi_{Bn}$ is the SBH for electrons, $\phi_m$ is the WF of the contact metal, $\chi$ is the electron affinity, and $b$ is the y-intercept. S is the pinning factor and can be calculated by $S = d\phi_{Bn}/d\phi_m$, where $S = 1$ corresponds to the Schottky-Mott rule in which the SBH is simply determined by the difference between $\phi_m$ and $\chi$ and $S = 0$ corresponds to absolute FL pinning, where the SBH is independent of the metal WF. The charge neutrality level, $\phi_{CNL}$, denotes the energy of the surface states at which the FL of the metal electrodes is pinned. The $\phi_{CNL}$ can be estimated by the following expression, which is derived from equation (2):

$$\phi_{CNL} = \frac{\chi + b}{1 - S}$$

(3)

Figure 3a compares the SBH as a function of the metal WF for the OML-embedded InSe devices with that of the control samples. The S and $\phi_{CNL}$ are calculated using the following parameters: $\chi_{InSe} = 4.6$ eV,[38] $\phi_{m,In} = 4.12$ eV, $\phi_{m,Ti} = 4.33$ eV, $\phi_{m,Cr} = 4.5$ eV, and $\phi_{m,Pd} = 5.12$ eV.[39] Remarkably, the InSe transistors with the OML-embedded tunneling contact manifest a large pinning factor of $S = 0.5 \pm 0.01$, indicating a pronounced FL depinning. This result is in contrast to that of the control samples, revealing a strong FL pinning with $S = 0.05 \pm 0.02$. Moreover, due to the FL depinning, the modulation of the SBH by varying the WF of the contact metals (0.5 eV) is more substantial than that of the control samples (0.05 eV). We note that the FL depinning in layered semiconductors with a tunneling contact has not been realized before, highlighting the unique contact properties of the metal/insulator/layered-semiconductor interface.

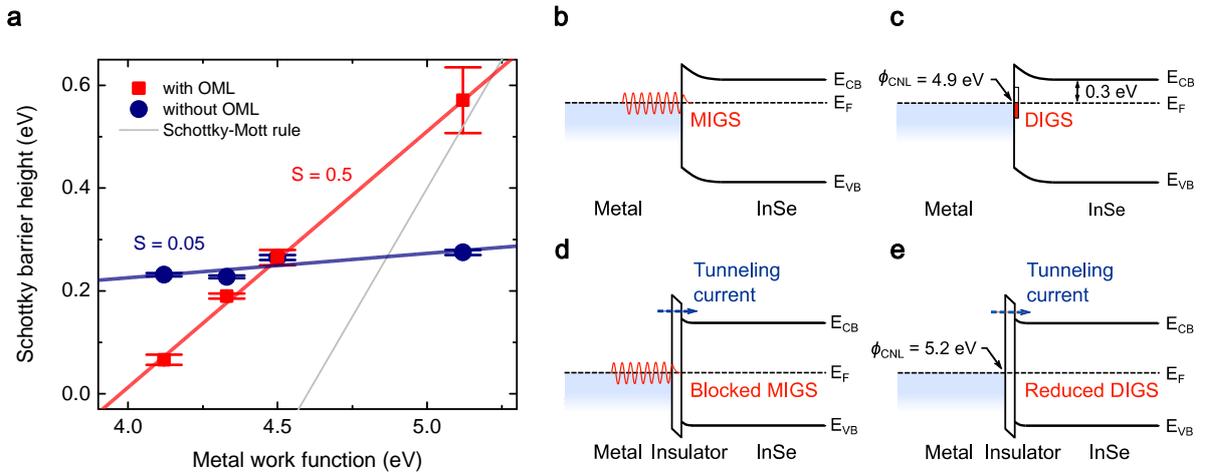

**Figure 3.** (a) The SBH as a function of metal WF for the OML-embedded InSe devices and the control samples. The InSe devices with the OML-embedded tunneling contact manifest a large pinning factor of S = 0.5, indicating a pronounced FL depinning. The gray line shows the Schottky-Mott rule in the case of no FL pinning effect. Schematics of energy band diagrams at the metal/InSe interfaces show (b) the MIGS resulting from the decaying metal wave function tailing into InSe, and (c) the DIGS resulting from defect states. With the insertion the OML, the tunneling barrier can facilitate (d) attenuation of the gap states resulting from the penetrated metal wave functions, and (e) mitigation of the DIGS in metal/InSe interfaces. The $\phi_{CNL}$ is estimated to be 5.2 eV and 4.9 eV with and without the OML, respectively.

The mechanism of the FL depinning in OML-embedded InSe transistors can be elucidated by examining the value of S and $\phi_{CNL}$. First, a theoretical model considering the MIGS model predicts a pinning factor of S = 0.2 in an ideal metal/InSe interface.[40, 41] The OML-embedded InSe devices show a greater S = 0.5, indicating that the presence of the OML can suppress MIGS, as illustrated in Figure 3b and 3d. Next, the $\phi_{CNL}$ is estimated by equation (3) to be 4.9 eV for the control samples. This $\phi_{CNL}$ is approximately 0.3 eV below the conduction band edge (CBE) ($\chi_{InSe}$ = 4.6 eV),[38] suggesting that the FL pinning in the control samples may be associated with defect states such as Se vacancies.[42] With the OML, the $\phi_{CNL}$ increases to 5.2 eV, signifying that the DIGS corresponding to $\phi_{CNL}$ = 4.9 eV are greatly suppressed, as depicted in Figure 3c and 3e. It is noteworthy that the FL depinning resulting from the tunneling contact can be mainly attributed to the inherently dangling-bond-free nature of the layered materials as well as the resist-free fabrication method of the metal contact. This clean interface and the high-quality InSe crystalline structure, as evidenced by the TEM characterizations, can mitigate the DIGS originating from the interfacial disorder. On the other hand, it has been reported that a tunneling barrier with a large band gap and small dielectric constant is beneficial to suppress FL pinning.[18] Whether using an alternative ultrathin oxide within the MIS contact of the layered semiconductor can lead to even more complete FL depinning is worth further investigation.

Enabled by the FL depinning, we now show that the InSe devices with the tunneling contact manifest highly tunable transport characteristics controlled by the WF of the contact metal. Figure 4a compares the semilogarithmic output curves for InSe devices with In ($\phi_{m,In} = 4.12$ eV) and Pd ($\phi_{m,Pd} = 5.12$ eV) contact metals. In the case of the In contact, the $I_{ds}$ of the OML-embedded InSe devices is increased by three-fold, indicating a reduced SBH. In contrast, the InSe devices with the Pd contact exhibit a decreased $I_{ds}$ and an increased contact barrier with the addition of the OML. This result is reasonable because for the Pd contact, we observe a larger SBH ($\phi_{Bn,Pd}$~0.57 eV) when the FL is not pinned compared with that of the pinning scenario ($\phi_{Bn,Pd}$~0.28 eV). On the other hand, the integration of the OML with the In contact leads to a small energy difference between the In WF and the CBE when the FL depinning occurs.[43]

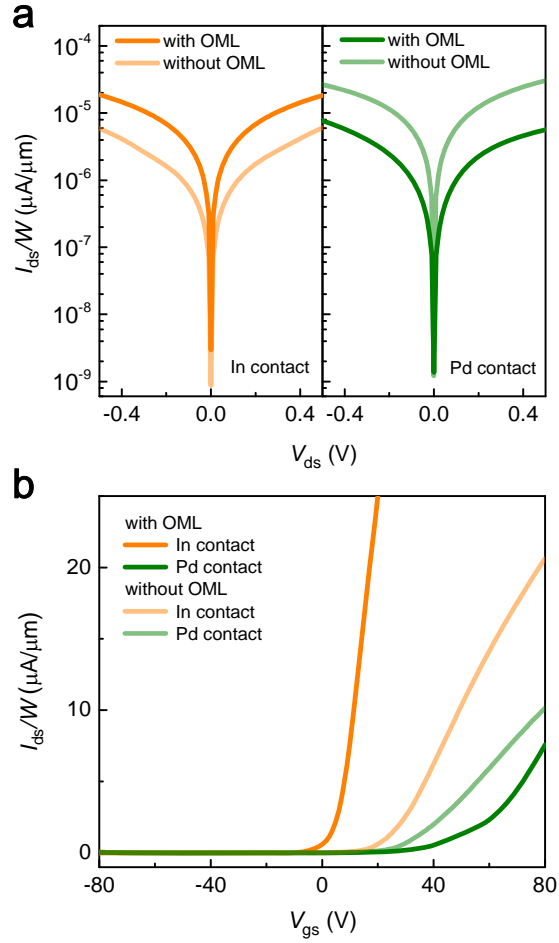

**Figure 4.** (a) Semilogarithmic output curves for the InSe devices with In metal contacts at $T = 2$ K and $V_{gs} = 80$ V. The $I_{ds}$ of InSe devices with the OML (orange line) is three-fold larger than that of InSe devices without the OML (light orange line), indicating a reduced SBH. Semilogarithmic output curves for the InSe devices with Pd metal contacts at $T = 2$ K and $V_{gs} = 80$ V. The InSe device with the OML (green line) exhibits a decreased $I_{ds}$ compared to that of the control sample (light green line). (b) The transfer curves of the InSe devices at $T = 2$ K with the OML (In contact: orange line; Pd contact: green line) and without the OML (In contact: light orange line; Pd contact: light green line). The applied source-drain voltage is 1 V. The changes in $V_{th}$ and $I_{ds}$ can be explained by the integration of the OML, further supporting the FL depinning in the InSe devices with the tunneling contact.

Figure 4b compares the transfer curves ($I_{ds} - V_{gs}$ curves) of the OML-embedded InSe devices and the control samples at $T = 2$ K. The control samples with In and Pd contact metals exhibit a comparable threshold voltage of $V_{th} = 30$ V, suggesting the presence of the FL pinning effect. Interestingly, $V_{th}$ shifts positively by 20 V and negatively by 25 V for larger (Pd) and smaller (In) WF metals, respectively. At low $T$, in the absence of the thermionic emission, $V_{th}$ is associated with the difference between the metal WF and the CBE of InSe. The effective modulation of $V_{th}$ by varying the WF of the contact metal thus confirms the occurrence of FL depinning in the InSe devices embedded with the OML. It is noted that the $V_{th}$ for the OML-embedded InSe samples with the In contact is approximately 0 V, which is reasonable because the energy difference between the In WF and the CBE is small.[35] On the other hand, owing to the high WF of Pd ($\phi_{m,Pd} = 5.12$ eV) relative to that of InSe ($\phi_{m,InSe} = 4.7$ eV), a large and positive $V_{gs}$ is needed to electrostatically gate the InSe channel to attain the turn-on state. The observed shifting in $V_{th}$ is not caused by the doping effect of the surface oxide,[44] as verified by OML-embedded InSe devices with an additional ozone treatment that causes further oxidation layer (Supplementary Fig. S9 and S10). Therefore, the observed changes in $V_{th}$ and $I_{ds}$ due to the integration of the OML further substantiate the FL depinning in the InSe devices with the tunneling contact.

We have presented that the InSe devices with the tunneling contact enable the effective modulation of the SBH by alleviating FL pinning. This tunneling contact scheme can thus be utilized to improve the device performance by lowering the contact resistance. Accordingly, we demonstrate high-quality InSe devices with the OML-embedded tunneling contact by employing the In contact, which exhibits a small SBH of 65 meV (Figure 3a). Figure 5a compares the transfer curves of an OML-embedded InSe device (sample A) and a control sample (sample B) at $T = 2$ K for $V_{ds} = 1$ V. Both samples exhibit an n-type conduction behavior; however, sample A shows a high turn-on current of 103 $\mu A/\mu m$ at $V_{gs} = 80$ V, which is approximately 6-times that of sample B. This enhanced $I_{ds}$ can be attributed to a large tunneling contact area as a result of the uniform oxide layer[45] and a low energy barrier corresponding to the tunneling contact. The inset of Figure 5a presents the logarithmic transfer curve of sample A, which exhibits a pronounced turn-on behavior. A large on/off ratio of $10^8$ at $T = 2$ K (and $10^7$ at $T = 300$ K) is observed, indicating an ideal transistor behavior.

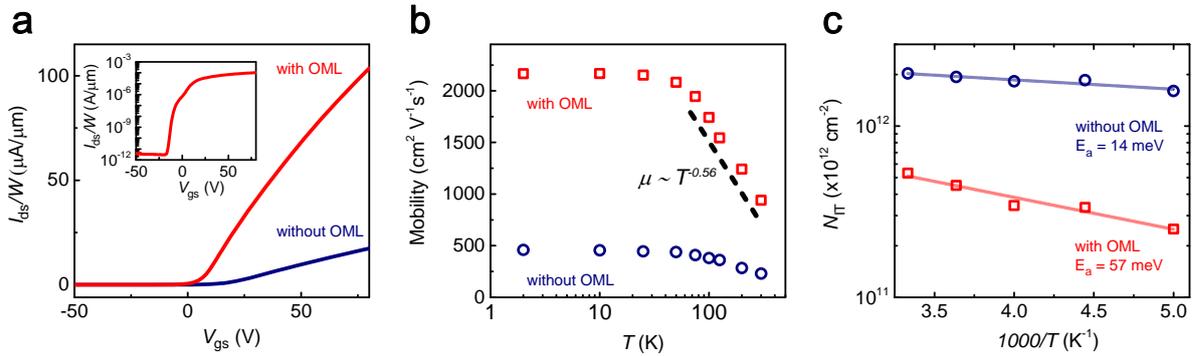

**Figure 5.** Electrical transport properties of the InSe devices. Comparisons of the (a) transfer characteristics and (b) two-terminal field-effect mobilities ($\mu_{FE}$) of the InSe devices with and without the OML-embedded tunneling contact. Inset: Semilogarithmic transfer characteristics of an InSe device with the OML, showing an ideal transistor behavior. The InSe transistor with the OML-embedded tunneling contact exhibits a high two-terminal $\mu_{FE}$ of 2160 cm$^2$/Vs and a large

on/off ratio of $10^8$ at $T = 2$ K. The source-drain voltage is 1 V. (c) A comparison of the surface trap densities of the InSe devices with and without the OML. The activation energies, $E_a$, of the InSe devices with and without the OML are extracted to be 57 and 14 meV, respectively.

Figure 5b compares the two-terminal field-effect mobility ($\mu_{FE}$) as a function of $T$ for samples A and B in which the $\mu_{FE}$ is calculated by $\mu_{FE} = (1/V_{ds}C_{SiO2})(L/W)(dI_{ds}/dV_{gs})$. Remarkably, sample A manifests a high two-terminal $\mu_{FE}$ of 2160 cm$^2$/Vs at $T = 2$ K and 940 cm$^2$/Vs at $T = 300$ K. This $\mu_{FE}$ compares favorably with other layered-semiconductor-based transistors with the tunneling contact.[31, 32, 46] The $\mu_{FE}$ of sample A is four-times that of the control sample, confirming the significance of introducing the OML into the metal/InSe interface. This enhancement in $\mu_{FE}$ can be attributed to the reduction in the contact barrier (Supplementary Fig. S7). At the high-$T$ regime ($T = 200 - 300$ K), the $T$ dependence of $\mu_{FE}$ can be described by $\mu_{ph} \propto T^{-\gamma}$ ($\gamma = 0.56$ and 0.52 for samples A and B, respectively), indicating a phonon-scattering-limited $\mu_{FE}$.

We further examine the trap states in our InSe devices to understand the effect of the OML on the carrier transport properties. The trap density in a transistor can be estimated by $N_{IT} = C_{ox}\Delta V_{th}/q$, where $C_{ox}$ is the gate capacitance of the h-BN/SiO$_2$ dielectric layer, $\Delta V_{th}$ is the shift in the $V_{th}$ between the forward and backward sweeps of the transfer curves, and $q$ is the elementary charge.[47] Figure 5c compares the trap density of the OML-embedded InSe devices with that of a control sample in an Arrhenius plot. We observe a larger trap density in the control sample, which may be attributed to the formation of a native oxide or the adsorption of moisture.[48, 49] In contrast, the InSe device with the OML exhibits a very low trap density of ~2 × 10$^{11}$ cm$^{-2}$, suggesting a higher interfacial quality. Moreover, the linear decrease in $N_{IT}$ with lowering $T$ in both devices is consistent with a thermally activated behavior.[50] The activation energy, $E_a$, of the InSe device with the OML is extracted to be 57 meV, suggesting that the trap states, albeit existing at a low density, negligibly affect the band-edge electrons due to the large energy difference between the trap states and the CBE. On the other hand, $E_a \sim 14$ meV is extracted for the control sample, indicating the presence of shallow traps. The small trap density and the large $E_a$ of the OML-embedded InSe devices are both advantageous to the carrier transport in the channel, further validating the prominent transport characteristics (Supplementary Fig. S11).

CONCLUSIONS

In this work, we demonstrate a pronounced FL depinning in InSe transistors by employing a tunneling contact realized by a high-quality OML. Considering the great interest in layered-semiconductor-based devices, control of the contact properties is especially important for device applications. We show that the unique characteristics of layered semiconductors can be utilized to form an atomically thin, ultra-uniform OML. This OML is integrated into the tunneling contact of InSe transistors, leading to a strong FL depinning with $S = 0.5$ and an effective modulation of the SBH. Accordingly, the low contact barrier is achieved with the tunneling contact and the In electrodes, enabling a high electron mobility of 2160 cm$^2$/Vs in two-terminal InSe devices. The realization of FL depinning in high-mobility InSe transistors embedded with the OML inspires a feasible tunneling contact scheme for 2D-semiconductor-based electronic and optoelectronic applications.

METHODS

**The InSe samples fabrication**

For TEM, AFM, and Raman spectroscopy characterizations, the InSe samples are mechanically exfoliated from InSe crystals onto a highly-doped Si substrate with a 300 nm $SiO_2$ dielectric layer under ambient conditions. After exfoliation, we grown the surface OL on InSe flakes by employing UV ozone treatment at 80 °C for 10 seconds. For electrical characterization, we fabricate back-gated devices with the contact structure illustrated in Figure 2a. The thickness of the InSe flakes used for the devices ranges from 10 to 20 nm, as confirmed by AFM. We exfoliate the InSe flakes onto polydimethylsiloxane (PDMS), which are then transferred onto pre-exfoliated h-BN flakes by a dry transfer technique. We employ the h-BN flakes as the substrate for the InSe channel to reduce the extrinsic carrier scattering sources at the bottom interface. The InSe/h-BN stacking structures are annealed in a furnace at 300 °C with mixed gas (95% Ar and 5% $H_2$) to improve the quality of the InSe/h-BN interface and to remove the residual polymer at the surfaces. These InSe/h-BN samples are then oxidized by the UV-ozone process at a temperature ranging from 80 to 100 °C for 10 seconds, followed by depositing electrical contacts (contact metal/Au of 15 nm/ 40 nm) with e-beam evaporation at a base pressure of $1 \times 10^{-7}$ Torr. To attain high uniformity of the tunneling barrier, we employ a resist-free method to define the source and drain contacts to avoid the potential resist residue in conventional lithography techniques, allowing us to study the FL depinning effect accurately.

**Measurement**

All the electrical characterization are performed in a Physical Property Measurement System in a helium environment at a pressure of 20 Torr. The channel conductance was measured by using a Keithley 2636 and the gate voltage was applied by using a Keithley 2400.

## Data availability

Data presented in this work is available upon reasonable request from the corresponding author.

REFERENCES

1. Wang, Q. H.; Kalantar-Zadeh, K.; Kis, A.; Coleman, J. N.; Strano, M. S., Electronics and optoelectronics of two-dimensional transition metal dichalcogenides. *Nat Nanotechnol* **2012,** *7* (11), 699-712.
2. Chhowalla, M.; Jena, D.; Zhang, H., Two-dimensional semiconductors for transistors. *Nat Rev Mater* **2016,** *1* (11), 16052.
3. Jariwala, D.; Sangwan, V. K.; Lauhon, L. J.; Marks, T. J.; Hersam, M. C., Emerging Device Applications for Semiconducting Two-Dimensional Transition Metal Dichalcogenides. *Acs Nano* **2014,** *8* (2), 1102-1120.
4. Sarkar, D.; Xie, X. J.; Liu, W.; Cao, W.; Kang, J. H.; Gong, Y. J.; Kraemer, S.; Ajayan, P. M.; Banerjee, K., A subthermionic tunnel field-effect transistor with an atomically thin channel. *Nature* **2015,** *526* (7571), 91-95.
5. Liu, Y. Y.; Stradins, P.; Wei, S. H., Van der Waals metal-semiconductor junction: Weak Fermi level pinning enables effective tuning of Schottky barrier. *Sci Adv* **2016,** *2* (4).
6. Liu, Y.; Guo, J.; Zhu, E. B.; Liao, L.; Lee, S. J.; Ding, M. N.; Shakir, I.; Gambin, V.; Huang, Y.; Duan, X. F., Approaching the Schottky-Mott limit in van der Waals metal-semiconductor junctions. *Nature* **2018,** *557* (7707), 696-+.
7. Kang, J. H.; Liu, W.; Sarkar, D.; Jena, D.; Banerjee, K., Computational Study of Metal Contacts to Monolayer Transition-Metal Dichalcogenide Semiconductors. *Phys Rev X* **2014,** *4* (3).
8. Tung, R. T., The physics and chemistry of the Schottky barrier height. *Appl Phys Rev* **2014,** *1* (1).
9. Allain, A.; Kang, J. H.; Banerjee, K.; Kis, A., Electrical contacts to two-dimensional semiconductors. *Nat Mater* **2015,** *14* (12), 1195-1205.
10. Das, S.; Chen, H. Y.; Penumatcha, A. V.; Appenzeller, J., High Performance Multilayer $MoS_2$ Transistors with Scandium Contacts. *Nano Lett.* **2013,** *13* (1), 100-105.
11. Tang, C. G.; Ang, M. C.; Choo, K.-K.; Keerthi, V.; Tan, J.-K.; Syafiqah, M. N.; Kugler, T.; Burroughes, J. H.; Png, R.-Q.; Chua, L.-L., Doped polymer semiconductors with ultrahigh and ultralow work functions for ohmic contacts. *Nature* **2016,** *539* (7630), 536.
12. Brillson, L. J.; Lu, Y., ZnO Schottky barriers and Ohmic contacts. *Journal of Applied Physics* **2011,** *109* (12), 8.
13. Gerling, L. G.; Mahato, S.; Morales-Vilches, A.; Masmitja, G.; Ortega, P.; Voz, C.; Alcubilla, R.; Puigdollers, J., Transition metal oxides as hole-selective contacts in silicon heterojunctions solar cells. *Solar Energy Materials and Solar Cells* **2016,** *145*, 109-115.
14. Kim, C.; Moon, I.; Lee, D.; Choi, M. S.; Ahmed, F.; Nam, S.; Cho, Y.; Shin, H. J.; Park, S.; Yoo, W. J., Fermi Level Pinning at Electrical Metal Contacts of Monolayer Molybdenum Dichalcogenides. *Acs Nano* **2017,** *11* (2), 1588-1596.
15. Park, W.; Kim, Y.; Lee, S. K.; Jung, U.; Yang, J. H.; Cho, C.; Kim, Y. J.; Lim, S. K.; Hwang, I. S.; Lee, B. H. In *Contact resistance reduction using Fermi level de-pinning layer for MoS 2 FETs*, 2014 IEEE International Electron Devices Meeting, IEEE: 2014; pp 5.1. 1-5.1. 4.
16. Bampoulis, P.; van Bremen, R.; Yao, Q. R.; Poelsema, B.; Zandvliet, H. J. W.; Sotthewes, K., Defect Dominated Charge Transport and Fermi Level Pinning in MoS2/Metal Contacts. *Acs Appl Mater Inter* **2017,** *9* (22), 19278-19286.
17. Zhou, Y.; Ogawa, M.; Han, X. H.; Wang, K. L., Alleviation of Fermi-level pinning effect on

metal/germanium interface by insertion of an ultrathin aluminum oxide. *Appl. Phys. Lett.* **2008,** *93* (20).
18. Kobayashi, M.; Kinoshita, A.; Saraswat, K.; Wong, H. S. P.; Nishi, Y., Fermi level depinning in metal/Ge Schottky junction for metal source/drain Ge metal-oxide-semiconductor field-effect-transistor application. *J. Appl. Phys.* **2009,** *105* (2).
19. Nishimura, T.; Kita, K.; Toriumi, A., A significant shift of Schottky barrier heights at strongly pinned metal/germanium interface by inserting an ultra-thin insulating film. *Appl Phys Express* **2008,** *1* (5).
20. Lee, S.; Tang, A.; Aloni, S.; Wong, H. S. P., Statistical Study on the Schottky Barrier Reduction of Tunneling Contacts to CVD Synthesized MoS2. *Nano Lett.* **2016,** *16* (1), 276-281.
21. Kim, G. S.; Kim, S. H.; Park, J.; Han, K. H.; Kim, J.; Yu, H. Y., Schottky Barrier Height Engineering for Electrical Contacts of Multilayered MoS2 Transistors with Reduction of Metal-Induced Gap States. *Acs Nano* **2018,** *12* (6), 6292-6300.
22. Mleczko, M. J.; Zhang, C. F.; Lee, H. R.; Kuo, H. H.; Magyari-Kope, B.; Moore, R. G.; Shen, Z. X.; Fisher, I. R.; Nishi, Y.; Pop, E., HfSe2 and ZrSe2: Two-dimensional semiconductors with native high-kappa oxides. *Sci Adv* **2017,** *3* (8).
23. Yamamoto, M.; Dutta, S.; Aikawa, S.; Nakaharai, S.; Wakabayashi, K.; Fuhrer, M. S.; Ueno, K.; Tsukagoshi, K., Self-Limiting Layer-by-Layer Oxidation of Atomically Thin WSe2. *Nano Lett.* **2015,** *15* (3), 2067-2073.
24. Bandurin, D. A.; Tyurnina, A. V.; Yu, G. L.; Mishchenko, A.; Zolyomi, V.; Morozov, S. V.; Kumar, R. K.; Gorbachev, R. V.; Kudrynskyi, Z. R.; Pezzini, S.; Kovalyuk, Z. D.; Zeitler, U.; Novoselov, K. S.; Patane, A.; Eaves, L.; Grigorieva, I. V.; Fal'ko, V. I.; Geim, A. K.; Cao, Y., High electron mobility, quantum Hall effect and anomalous optical response in atomically thin InSe. *Nat Nanotechnol* **2017,** *12* (3), 223-227.
25. Feng, W.; Zheng, W.; Cao, W.; Hu, P., Back gated multilayer InSe transistors with enhanced carrier mobilities via the suppression of carrier scattering from a dielectric interface. *Adv Mater* **2014,** *26* (38), 6587-93.
26. Mudd, G. W.; Svatek, S. A.; Ren, T.; Patane, A.; Makarovsky, O.; Eaves, L.; Beton, P. H.; Kovalyuk, Z. D.; Lashkarev, G. V.; Kudrynskyi, Z. R.; Dmitriev, A. I., Tuning the Bandgap of Exfoliated InSe Nanosheets by Quantum Confinement. *Adv. Mater.* **2013,** *25* (40), 5714-5718.
27. Chang, Y. R.; Ho, P. H.; Wen, C. Y.; Chen, T. P.; Li, S. S.; Wang, J. Y.; Li, M. K.; Tsai, C. A.; Sankar, R.; Wang, W. H.; Chiu, P. W.; Chou, F. C.; Chen, C. W., Surface Oxidation Doping to Enhance Photogenerated Carrier Separation Efficiency for Ultrahigh Gain Indium Selenide Photodetector. *Acs Photonics* **2017,** *4* (11), 2930-2936.
28. DeRosa, M. C.; Crutchley, R. J., Photosensitized singlet oxygen and its applications. *Coord. Chem. Rev.* **2002,** *233*, 351-371.
29. Yamamoto, M.; Einstein, T. L.; Fuhrer, M. S.; Cullen, W. G., Anisotropic Etching of Atomically Thin MoS2. *Journal of Physical Chemistry C* **2013,** *117* (48), 25643-25649.
30. Dean, C. R.; Young, A. F.; Meric, I.; Lee, C.; Wang, L.; Sorgenfrei, S.; Watanabe, K.; Taniguchi, T.; Kim, P.; Shepard, K. L.; Hone, J., Boron nitride substrates for high-quality graphene electronics. *Nat Nanotechnol* **2010,** *5* (10), 722-726.
31. Wang, J. L.; Yao, Q.; Huang, C. W.; Zou, X. M.; Liao, L.; Chen, S. S.; Fan, Z. Y.; Zhang, K.; Wu, W.; Xiao, X. H.; Jiang, C. Z.; Wu, W. W., High Mobility MoS2 Transistor with Low Schottky Barrier Contact by Using Atomic Thick h-BN as a Tunneling Layer. *Adv. Mater.* **2016,** *28* (37), 8302-

8308.
32. Cui, X.; Shih, E. M.; Jauregui, L. A.; Chae, S. H.; Kim, Y. D.; Li, B. C.; Seo, D.; Pistunova, K.; Yin, J.; Park, J. H.; Choi, H. J.; Lee, Y. H.; Watanabe, K.; Taniguchi, T.; Kim, P.; Dean, C. R.; Hone, J. C., Low-Temperature Ohmic Contact to Monolayer MoS2 by van der Waals Bonded Co/h-BN Electrodes. *Nano Lett.* **2017,** *17* (8), 4781-4786.
33. Nan, H. Y.; Guo, S. J.; Cai, S.; Chen, Z. R.; Zafar, A.; Zhang, X. M.; Gu, X. F.; Xiao, S. Q.; Ni, Z. H., Producing air-stable InSe nanosheet through mild oxygen plasma treatment. *Semicond. Sci. Technol.* **2018,** *33* (7).
34. Sanchez-Royo, J. F.; Munoz-Matutano, G.; Brotons-Gisbert, M.; Martinez-Pastor, J. P.; Segura, A.; Cantarero, A.; Mata, R.; Canet-Ferrer, J.; Tobias, G.; Canadell, E.; Marques-Hueso, J.; Gerardot, B. D., Electronic structure, optical properties, and lattice dynamics in atomically thin indium selenide flakes. *Nano Research* **2014,** *7* (10), 1556-1568.
35. Huang, Y. T.; Chen, Y. H.; Ho, Y. J.; Huang, S. W.; Chang, Y. R.; Watanabe, K.; Taniguchi, T.; Chiu, H. C.; Liang, C. T.; Sankar, R.; Chou, F. C.; Chen, C. W.; Wang, W. H., High-Performance InSe Transistors with Ohmic Contact Enabled by Nonrectifying Barrier-Type Indium Electrodes. *Acs Appl Mater Inter* **2018,** *10* (39), 33450-33456.
36. Chen, S. Y.; Ho, P. H.; Shiue, R. J.; Chen, C. W.; Wang, W. H., Transport/Magnetotransport of High-Performance Graphene Transistors on Organic Molecule-Functionalized Substrates. *Nano Lett.* **2012,** *12* (2), 964-969.
37. Robertson, J., Band offsets, Schottky barrier heights, and their effects on electronic devices. *J Vac Sci Technol A* **2013,** *31* (5).
38. Kudrynskyi, Z. R.; Bhuiyan, M. A.; Makarovsky, O.; Greener, J. D. G.; Vdovin, E. E.; Kovalyuk, Z. D.; Cao, Y.; Mishchenko, A.; Novoselov, K. S.; Beton, P. H.; Eaves, L.; Patane, A., Giant Quantum Hall Plateau in Graphene Coupled to an InSe van der Waals Crystal. *Phys. Rev. Lett.* **2017,** *119* (15), 157701.
39. Michaelson, H. B., The work function of the elements and its periodicity. *Journal of applied physics* **1977,** *48* (11), 4729-4733.
40. Brudnyi, V. N.; Sarkisov, S. Y.; Kosobutsky, A. V., Electronic properties of GaSe, InSe, GaS and GaTe layered semiconductors: charge neutrality level and interface barrier heights. *Semicond. Sci. Technol.* **2015,** *30* (11).
41. Guo, Y. Z.; Robertson, J., Band structure, band offsets, substitutional doping, and Schottky barriers of bulk and monolayer InSe. *Phys Rev Mater* **2017,** *1* (4).
42. Kistanov, A. A.; Cai, Y.; Zhou, K.; Dmitriev, S. V.; Zhang, Y. W., Atomic-scale mechanisms of defect- and light-induced oxidation and degradation of InSe. *J Mater Chem C* **2018,** *6* (3), 518-525.
43. Zheng, S.; Lu, H. C.; Liu, H.; Liu, D. M.; Robertson, J., Insertion of an ultrathin Al2O3 interfacial layer for Schottky barrier height reduction in WS2 field-effect transistors. *Nanoscale* **2019,** *11* (11), 4811-4821.
44. Cai, L. L.; McClellan, C. J.; Koh, A. L.; Li, H.; Yalon, E.; Pop, E.; Zheng, X. L., Rapid Flame Synthesis of Atomically Thin MoO3 down to Monolayer Thickness for Effective Hole Doping of WSe2. *Nano Lett.* **2017,** *17* (6), 3854-3861.
45. Giannazzo, F.; Fisichella, G.; Greco, G.; Di Franco, S.; Deretzis, I.; La Magna, A.; Bongiorno, C.; Nicotra, G.; Spinella, C.; Scopelliti, M.; Pignataro, B.; Agnello, S.; Roccaforte, F., Ambipolar MoS2 Transistors by Nanoscale Tailoring of Schottky Barrier Using Oxygen Plasma

Functionalization. *Acs Appl Mater Inter* **2017,** *9* (27), 23164-23174.
46. Avsar, A.; Tan, J. Y.; Luo, X.; Khoo, K. H.; Yeo, Y. T.; Watanabe, K.; Taniguchi, T.; Quek, S. Y.; Ozyilmaz, B., van der Waals Bonded Co/h-BN Contacts to Ultrathin Black Phosphorus Devices. *Nano Lett.* **2017,** *17* (9), 5361-5367.
47. Yamamoto, M.; Ueno, K.; Tsukagoshi, K., Pronounced photogating effect in atomically thin WSe2 with a self-limiting surface oxide layer. *Appl. Phys. Lett.* **2018,** *112* (18).
48. Ho, P. H.; Chang, Y. R.; Chu, Y. C.; Li, M. K.; Tsai, C. A.; Wang, W. H.; Ho, C. H.; Chen, C. W.; Chiu, P. W., High-Mobility InSe Transistors: The Role of Surface Oxides. *Acs Nano* **2017,** *11* (7), 7362-7370.
49. Late, D. J.; Liu, B.; Matte, H. S. S. R.; Dravid, V. P.; Rao, C. N. R., Hysteresis in Single-Layer MoS2 Field Effect Transistors. *Acs Nano* **2012,** *6* (6), 5635-5641.
50. Kordos, P.; Donoval, D.; Florovic, M.; Kovac, J.; Gregusova, D., Investigation of trap effects in AlGaN/GaN field-effect transistors by temperature dependent threshold voltage analysis. *Appl. Phys. Lett.* **2008,** *92* (15).

**Supplementary Information**.

X-ray photoelectron spectroscopy (XPS) study. The AFM characterization of bulk h-BN and the InSe without UV-ozone treatment. Substrate effect on the transport properties. Tunneling current characteristics enabled by the metal/insulator/semiconductor (MIS) contact with the OL as the tunneling barrier. Improved transport characteristics and contact of the OML-embedded InSe devices. Extraction of the Schottky barrier height (SBH) of the InSe devices. The OML-embedded InSe devices with an additional UV-ozone treatment. The distribution of the OML-embedded InSe devices and the control samples as a function of room-$T$ $\mu_{FE}$ and $N_{IT}$.

**Corresponding Author**

*(W.-H. Wang) Tel: +886-2-2366-8208, Fax: +886-2-2362-0200, wwang@sinica.edu.tw

**Author Contributions**

Y. H. Chen, C. Y. Cheng, and W. H. Wang conceived the research and designed the experiment. Y. H. Chen performed device fabrication, characterizations, and data analysis. S. Y. Chen contributed to the Raman spectroscopy measurement and analysis. J. S. D. Rodriguez and H. S. Chiu performed the AFM measurement and analysis. H. T. Liao contributed the Raman spectroscopy measurement. K. Watanabe and T. Taniguchi contributed h-BN material. R. Sankar and F. C. Chou contributed InSe material. All authors discussed the results and commented on the manuscript. Y. H. Chen, C. Y. Cheng, S. Y. Chen, and W. H. Wang wrote the manuscript.

**Competing interests**

The authors declare that there are no competing interests.

**Acknowledgment**

This work was supported by the Ministry of Science and Technology of Taiwan, R.O.C. under contract numbers MOST 103-2112-M-001-020-MY3.

# Oxidized-monolayer Tunneling Barrier for Strong Fermi-level Depinning in layered InSe Transistors

*Yi-Hsun Chen[1], Chih-Yi Cheng[1], Shao-Yu Chen[1,2], Jan Sebastian Dominic Rodriguez[3], Han-Ting Liao[4], Kenji Watanabe[5], Takashi Taniguchi[5], Chun-Wei Chen[4], Raman Sankar[6,7], Fang-Cheng Chou[6], Hsiang-Chih Chiu[3] and Wei-Hua Wang[1]\**

[1]Institute of Atomic and Molecular Sciences, Academia Sinica, Taipei 106, Taiwan

[2]School of Physics and Astronomy, Monash University, Victoria 3800, Australia

[3]Department of Physics, National Normal Taiwan University, Taipei 106, Taiwan

[4]Department of Materials Science and Engineering, National Taiwan University, Taipei 106, Taiwan

[5]Advanced Materials Laboratory, National Institute for Materials Science, Tsukuba, 305-0044, Japan

[6]Center of Condensed Matter Sciences, National Taiwan University, Taipei 106, Taiwan

[7]Institute of Physics, Academia Sinica, Taipei 115, Taiwan

## S1. X-ray photoelectron spectroscopy (XPS) study

We analyze the elemental and chemical composition of the surface-oxidized InSe by employing X-ray photoelectron spectroscopy (XPS) measurements. To avoid the formation of a native oxide, the InSe flakes were transferred to the ultrahigh vacuum chamber of the XPS system right after the mechanical exfoliation. Further discussion regarding native oxide is presented in Supplementary Fig. S2, S3, and S4. To understand the effect of surface oxidation, we compare the XPS signals of OML-embedded InSe and pristine InSe samples, as shown in Supplementary Figure S1a and S1b. Supplementary Figure S1a shows the XPS spectrum of pristine InSe in which the doublet peaks at 445.3 and 452.9 eV can be attributed to the In $3d_{5/2}$ and In $3d_{3/2}$ core levels, respectively.[1] For InSe treated with UV ozone, the XPS peaks become broadened, and two small doublet peaks appear at the high-energy side of the In $3d_{5/2}$ and In $3d_{3/2}$ core levels. The binding energy difference of the two peaks is 0.9 eV, suggesting the presence of $In_2O_3$ and that $In^{+3}$ was oxidized from $In^{+2}$.[2] Moreover, the core levels of In $3d_{5/2}$ and In $3d_{3/2}$ exhibit a redshift to 445.2 and 452.7 eV, respectively, which can be attributed to the p-type doping caused by the oxidation layer.[3]

Supplementary Figure S1b compares the Se 3d core levels of the InSe sample treated with UV ozone with that of the pristine InSe sample. For pristine InSe, the doublet peaks at 54.8 and 55.6 eV correspond to the Se $3d_{5/2}$ and $3d_{3/2}$ core levels, respectively.[1] For InSe with the UV-ozone treatment, the $3d_{5/2}$ and $3d_{3/2}$ lines show a redshift while retaining the energy difference between the $3d_{5/2}$ and $3d_{3/2}$ lines, suggesting pristine InSe is underneath the OML. Moreover, doublet peaks with binding energies of 59.2 and 60.0 eV appear, indicating the transformation from $Se^{-2}$ to $Se^{+4}$ and the presence of nonstoichiometric $SeO_x$ with x~2 after the UV-ozone treatment.[4,5] In addition, the doublet peaks for the Se $3d_{5/2}$ and $3d_{3/2}$ core levels exhibit a redshift and a broadening, showing a similar trend as that of the In 3d core levels. Therefore, the XPS study confirms the presence of oxidized species in the UV-ozone treated InSe and reveals the chemical composition of the OML.

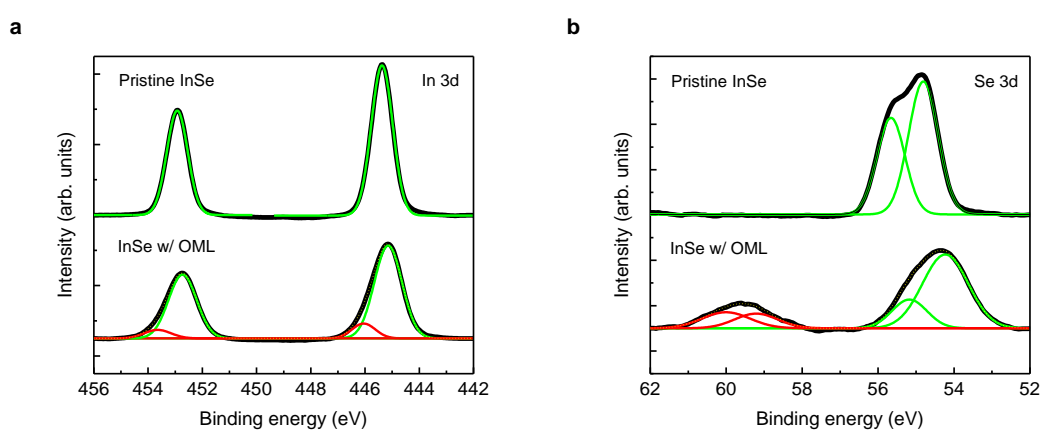

**Supplementary Figure S1**. The XPS spectra of InSe with the OML and the control InSe sample. (a) In 3d core level spectra and (b) Se 3d core level spectra.

We further compare the XPS spectra of the InSe sample oxidized by UV ozone with that consists of the native oxide (**Supplementary Figure S2**) to assess the chemical composition. The InSe sample with the native oxide was kept in air (40% humidity and 25°C) for 120 hours, and the XPS data is adapted from our previous study [4]. As aforementioned, the UV-ozone-treated InSe sample exhibits additional doublet peaks which appear at the high-energy side of the In $3d$ core level (the difference of peak energy is 0.9 eV), suggesting the presence of $In_2O_3$ and that $In^{+3}$ was oxidized from $In^{+2}$ [5, 6]. For the InSe sample with native oxide, additional XPS peak also appears at the high-energy side of the In $3d$ core level. However, the binding energy difference of the two peaks ranges from 0.3 to 0.7 eV, suggesting a more complex form of oxide with In other than $In^{+3}$. For Se $3d$ core level spectra, doublet peaks at the high-energy side appear for both samples, suggesting the presence of nonstoichiometric $SeO_x$ with x~2.

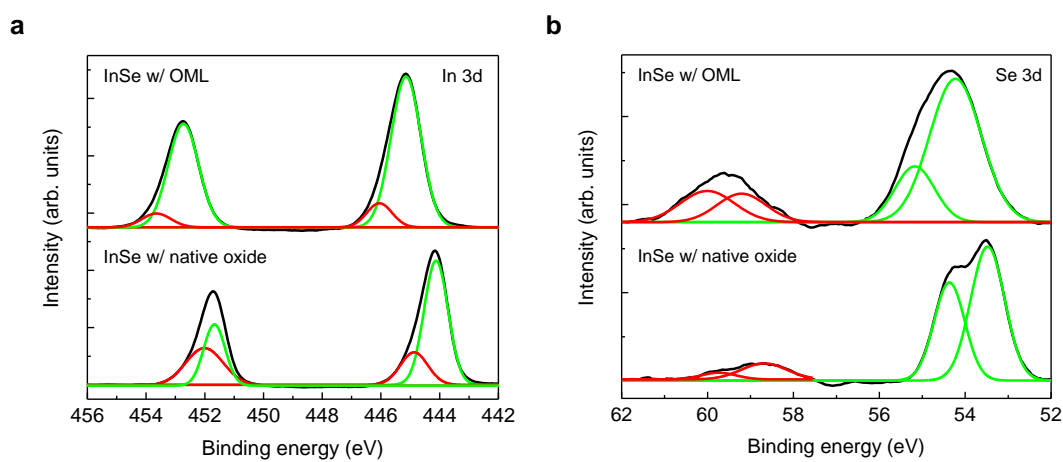

**Supplementary Figure S2**. A comparison of XPS spectra of UV-ozone-treated InSe with the InSe consists of native oxide. (a) In 3d core level spectra and (b) Se 3d core

level spectra. The XPS data for the InSe sample with the native oxide is adapted from our previous study [4].

## S2. The AFM characterization of bulk h-BN and the InSe without UV-ozone treatment

Supplementary Figure S3a and S3b shows the surface morphology of bulk h-BN and the InSe sample without the UV-ozone treatment (control InSe sample), respectively. The AFM images are acquired by a Bruker Multimode 8 AFM with Peak-Force Quantitative Nanoscale Mechanical Characterization (PF-QNM). The scan window is $500 \times 500$ nm$^2$ with a resolution of 256 sample/line. The h-BN surface is atomic-scale smooth over a large area, consistent with previous studies.[6] In comparison with the InSe with the OML, the surface of the control InSe sample exhibits nucleation sites with a height $< 2$ nm (Supplementary Figure S3b and S3c), which can be attributed to the formation of the native oxide.[1] By fitting the histogram of the surface with a Gaussian function, we can estimate the surface roughness as the standard deviation. In the area without the nucleation sites, the surface roughness is 62 pm.

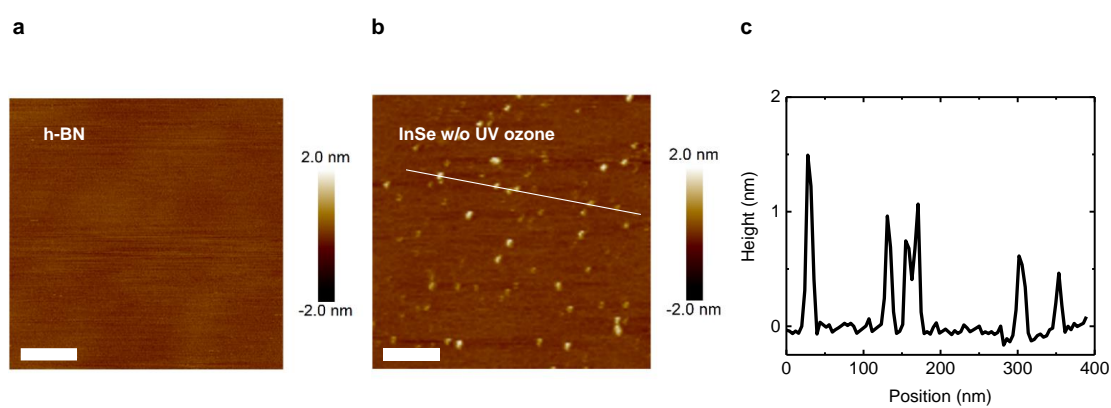

**Supplementary Figure S3**. The AFM images of (a) bulk h-BN and (b) the InSe sample

without the UV-ozone treatment. The scale bar is 100 nm. (c) The height profile corresponds to the white solid line in (b).

We performed an AFM characterization of the natively oxidized InSe to monitor the height and distribution of the native oxides under the ambient condition (40% humidity and 25°C) at various oxidation stages. Supplementary Figure S4 shows the AFM images of a control InSe sample taken at 15 min, 30 min, and 6.5 hrs after the mechanical exfoliation. As can be seen, the native oxides are still not observable at 15 min after the exfoliation (Figure S4a). A large number of the native oxides appear at nucleation sites at 30 min after the exfoliation (Figure S4b). After exposure to air for 6.5 hrs, the InSe surface shows negligible change (Figure S4c) compared with the surface topography at 30 min, indicating that the oxidation process has stopped. This observation clearly indicates that the native oxides would be improbable to form a uniform oxide layer, in contrast with UV-ozone-generated oxide in our devices, and the two oxidation process are completely different.

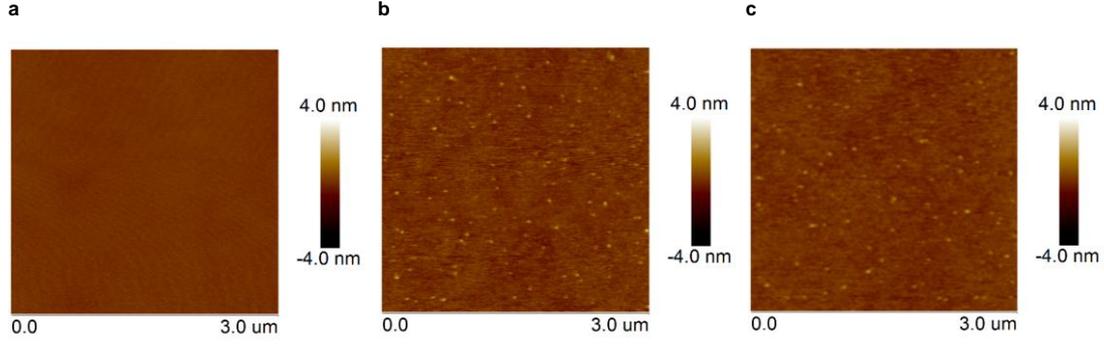

**Supplementary Figure S4**. AFM images of air-oxidized InSe, measured at (a) 15 minutes, (b) 30 minutes, and (c) 6 hours 30 minutes after the mechanical exfoliation of the InSe control sample.

We evaluate the contact area between the sharp tip and the InSe surface by using Hertz contact mechanics:

$$A = \pi \times \left(\frac{3}{4} \times \frac{R}{E'} \times (F_N + F_{Adhesion})\right)^{2/3} \quad (1)$$

$$\frac{1}{E'} = \left(\frac{1-v_{tip}^2}{E_{tip}} + \frac{1-v_{InSe}^2}{E_{InSe}}\right) \quad (2)$$

where $R$ is the radius of the sharp tip; $F_N$ is the normal force; $F_N$ is the adhesive force; $E'$ is the effective Young's modulus; $E_{tip}$ and $E_{InSe}$ are the Young's moduli of the sharp tip and InSe surface, respectively; and $v_{tip}$ and $v_{InSe}$ are the Poisson ratios of the sharp tip and the InSe surface, respectively. The $E_{InSe}$ and $v_{InSe}$ of InSe are 23.1 GPa and 0.27, respectively.[7] The contact area of the sharp tip on the InSe surface is then calculated by using equations (1) and (2). Supplementary Table 1 summarizes the contact area, the Young's modulus, and Poisson ratio of InSe. By taking the square root of the contact area, the spatial resolution is estimated to be 1 nm, highlighting an ultrahigh resolution by using a sharp tip and by operating in the Peak-Force mode.

**Supplementary Table 1**. The calculation of the contact area between the sharp tip and InSe surface

| Contact area (nm$^2$) | Young's modulus (GPa) | | Poisson ratio | |
|---|---|---|---|---|
| | Tip | InSe | Tip | InSe |
| 1.098 | 165 | 23.1 | 0.2 | 0.27 |

## S3. Substrate effect on the transport properties

We discuss the substrate effect on the electrical transport properties of the InSe devices with the OML. Supplementary Figure S5a shows the transfer curves of OML-embedded InSe devices fabricated on h-BN and SiO$_2$/Si substrates at $T = 300$ and $2$ K. Both devices exhibit an n-type conduction behavior. However, the InSe device fabricated on h-BN exhibits a high on-current compared with that of the InSe device fabricated on the SiO$_2$/Si substrate at both $T = 2$ K and $T = 300$ K. The applied source-drain voltage is 1 V. Supplementary Figure S5b compares the $\mu_{FE}$ for the InSe devices with the OML fabricated on the h-BN and SiO$_2$/Si substrates. The $\mu_{FE}$ of the InSe device fabricated on h-BN at $T = 2$ K is approximately 5-times higher than that of the InSe device fabricated on the SiO$_2$/Si substrate, signifying a reduction in the extrinsic scattering in the InSe channel.

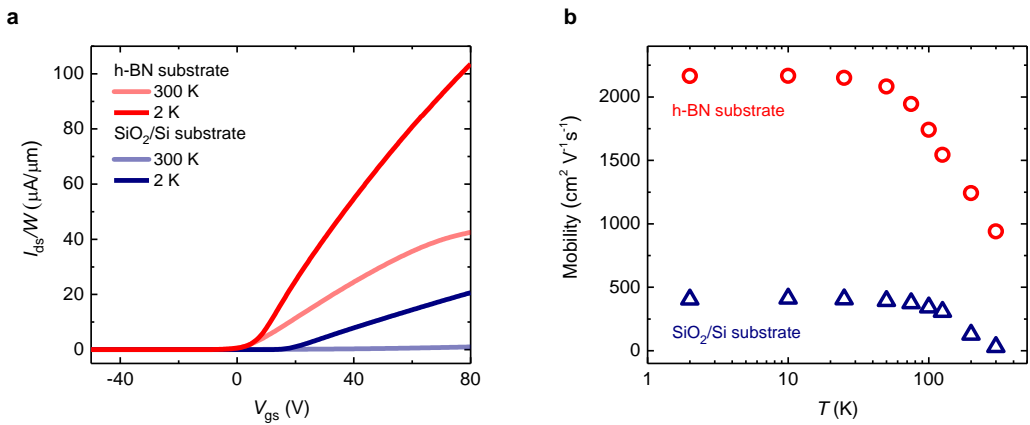

**Supplementary Figure S5**. Comparison of (a) the transfer characteristics at $T = 300$ and $2$ K, and (b) the two-terminal field-effect mobility ($\mu_{FE}$) for the OML-embedded

InSe devices fabricated on the h-BN (red) and SiO$_2$/Si substrates (blue). The source-drain voltage is 1 V.

## S4. Tunneling current characteristics enabled by the metal/insulator/semiconductor (MIS) contact with the OL as the tunneling barrier

Simmon's model is employed to describe the output curves of the InSe devices with the MIS contact and to examine the tunneling current behavior. According to Simmon's model, the current density in a symmetric tunneling junction can be described as follows[8,9]:

$$J = J_0 \varphi^{\frac{1}{2}} exp\left(-A\varphi^{\frac{1}{2}}\right) \qquad \text{for low-voltage range (V~0),}$$

$$J = J_0(V + \sigma V^3) \qquad \text{for intermediate-voltage range (V < } \varphi/e\text{),}$$

where

$$J_0 = \left[(2m_e)^{\frac{1}{2}}/d\right](e/h)^2, \tag{3}$$

$$A = (4\pi\beta d/h)(2m_e)^{\frac{1}{2}}, \tag{4}$$

$$\sigma = [(Ae)^2/96\varphi] - \left[Ae^2/32\varphi^{\frac{3}{2}}\right], \tag{5}$$

$$\beta = 1 - \frac{1}{8\bar{f}^2 d}\int_{d_1}^{d_2}[f(x) - \bar{f}]^2 dx, \tag{6}$$

where $m_e$ is the effective mass of an electron, $h$ is Planck's constant, $e$ is the elementary charge, $\beta$ is an arbitrary function, $d$ is the thickness of the tunneling barrier, and $\varphi$ is the height of the tunneling barrier. Supplementary Figure S6a-c show the output curves of the InSe devices utilizing UV-ozone processing temperatures of 80, 90, and 100 °C, respectively, at $T = 2$ K. Simmon's model can reasonably describe these output curves, which show a linear behavior at a very small $V_{ds}$ range (~0–0.05 V), followed by the cubic of $V_{ds}$ ranging from 0.05 to 0.15 V. We can therefore associate the $V_{ds}$ range of 0 to 0.05 V with the low-voltage regime and 0.05 to 0.15 V with the intermediate-voltage regime.

In the calculation, the tunneling barrier is assumed to be rectangular, that is, $\beta = 1$. The thickness of the tunneling barrier is 1 nm, in accordance with the TEM image shown in Figure 2a in the main text. The height of the tunneling barrier is estimated to be 6 meV by fitting the output curve of the OML-embedded InSe device utilizing a UV-ozone processing temperature of 80 °C at $T = 2$ K. We then use this barrier height to fit the output curves for the processing temperatures of 90 and 100°C by assuming that the chemical composition of the OL does not vary with the processing temperature and that only the thickness of the OL increases with the processing temperature. Supplementary Figure S6d shows the barrier thickness as a function of $\sigma$ by using equation (5) with a barrier height of 6 meV. The comparison of $\sigma$ extracted from Supplementary Figure S6b and S6c with the curve leads to an estimation of the thickness of the OL to be 1.15 nm and 1.8 nm for 90 and 100 °C, respectively. The analysis of the output curves by Simmons' model suggests that the tunneling effect governs the charge transport in the OL-embedded InSe devices.

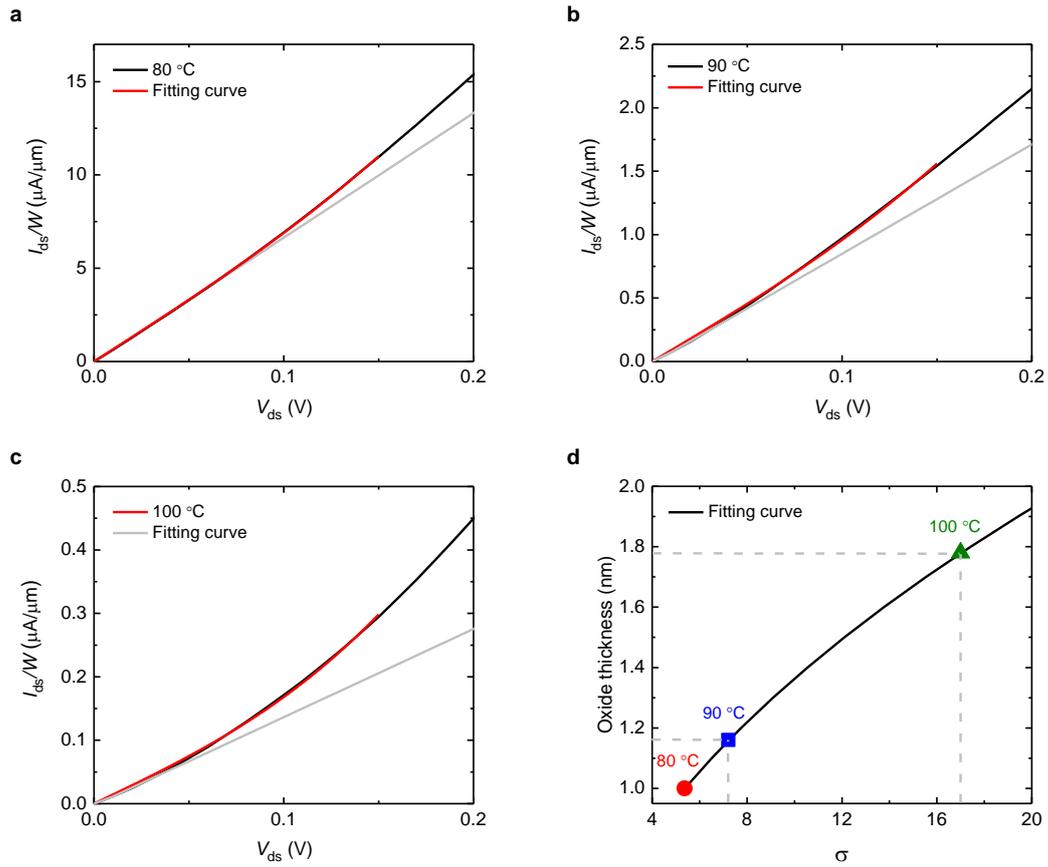

**Supplementary Figure S6**. The output curves of the OL-embedded InSe devices utilizing UV-ozone process temperatures of (a) 80, (b) 90, and (c) 100 °C, at $T = 2$ K and $V_{gs} = 80$ V. (d) The estimated thickness of the tunneling barrier as a function of $\sigma$.

## S5. Improved transport characteristics and contact of the OML-embedded InSe devices

We show additional transport characteristics of sample A (InSe device with the OML) and sample B (InSe device without the OML) discussed in Figure 5 in the main text. Supplementary Figure S7a shows the $T$-dependent output ($I_{ds} - V_{ds}$) curve of sample A. The output curves of sample A are linear and symmetric at high $T$ and become nonlinear at low $T$ (e.g., $T = 2$ K). The nonlinear characteristics at low $T$ suggests a typical tunneling current behavior corresponding to the MIS contact. As $T$ increases, the thermionic emission current that can overcome the energy barrier corresponding to the OML arises and ultimately dominates the carrier transport, leading to the linear and symmetric output curves. In contrast, sample B exhibits nonlinear curves at all $T$ (Supplementary Figure S7b), suggesting that the Schottky contact dominates at the metal/InSe interface.

Supplementary Figure S7c and S7d shows the output curves of samples A and B, respectively, at $T = 2$ K with $V_{gs}$ ranging from 0 to 80 V. Sample A manifests a high current density of 100 µA/µm at $V_{ds} = 1$ V and $V_{gs} = 80$ V compared with that of sample B (30 µA/µm), indicating the improved contact condition due to the insertion of the OML. The output curves of sample B show a nonlinear behavior, suggesting the presence of a large Schottky barrier at the metal/InSe interface. To further examine the contact barrier height, the Schottky barrier height (SBH) values of sample A and sample B are extracted by calculating $E_A$ with the thermionic emission equation. The SBHs extracted under the flat-band condition are 53 meV and 235 meV for sample A and sample B, respectively. The enhanced current density and reduced barrier height shown here further support the tunneling contact behavior as a result of inserting the OML.[10]

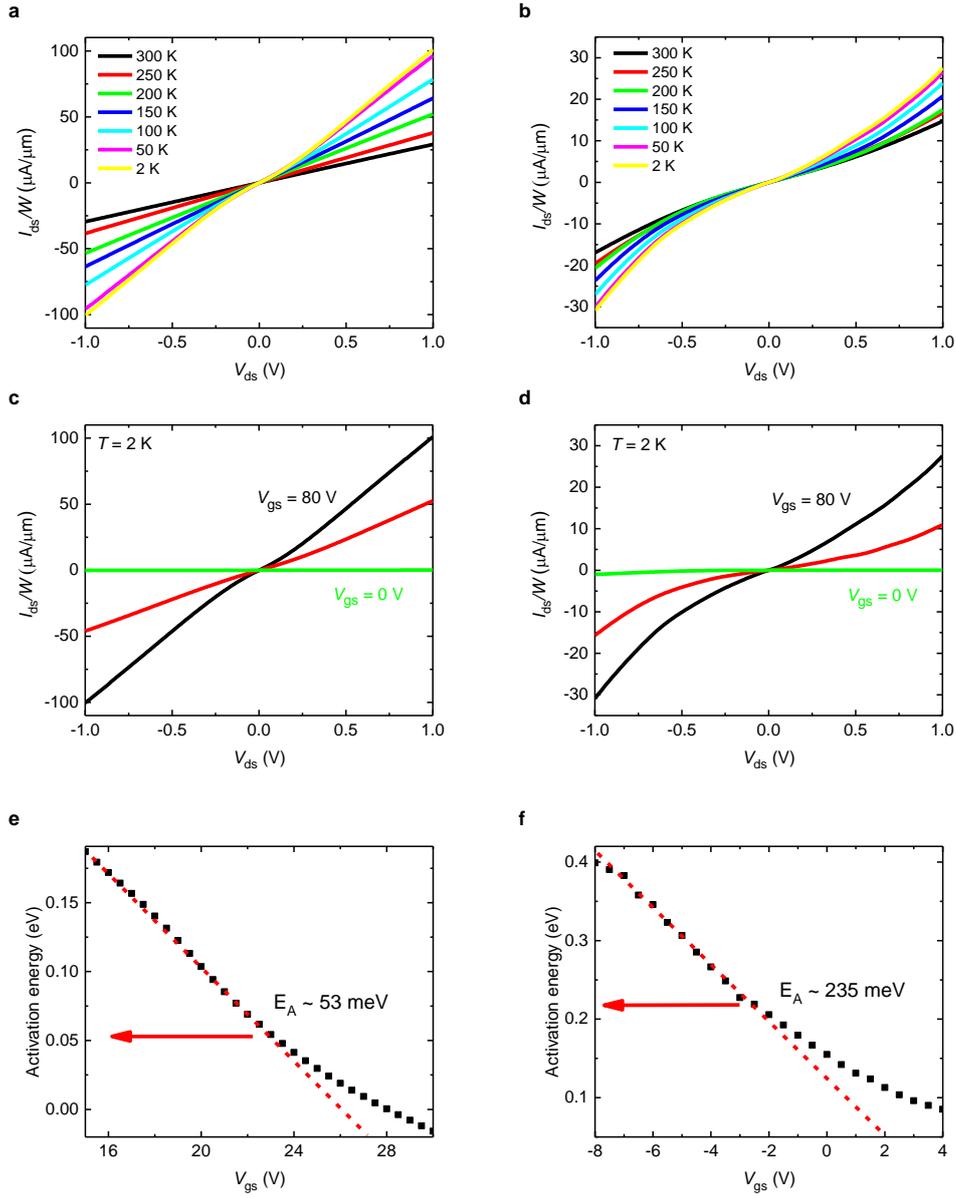

**Supplementary Figure S7.** $T$-dependent output curves of the InSe devices (a) with and (b) without the OML at $V_{gs} = 80$ V with $T$ ranging from 300 to 2 K. Output curves of the InSe devices (c) with and (d) without the OML at $T = 2$ K with $V_{gs}$ ranging from 0 to 80 V. Extracted SBHs as a function of $V_{gs}$ for the InSe devices (e) with and (f) without the OML.

## S6. Extraction of the Schottky barrier height (SBH) of the InSe devices

As discussed here, the SBH shown in Figure 3b in the main text is extracted by using the thermionic emission model. Supplementary Figure S8a, S8b, S8c, and S8d shows the Arrhenius plots of $I_{ds}/T^{3/2}$ for the OML-embedded InSe devices with In, Ti, Cr, and Pd contacts, respectively. The data can be well described by a linear fit in the $T$ range of $T = 300 - 400$ K. From Supplementary Figure S8a-d, we can calculate the $V_{gs}$ dependence of the activation energy, $E_A$, by using equation (1) in the main text for the corresponding devices, as shown in Supplementary Figure S8e-h. $E_A$ varies linearly with $V_{gs}$ when the thermionic emission dominates the drain current. On the other hand, when the tunneling current arises in addition to the thermionic emission current as a result of the shortening barrier, $E_A$ will deviate from linearity. At this turning point, the flat-band condition in which the band bending vanishes can be determined, and $E_A$ and the $V_{gs}$ represent the SBH and the $V_{FB}$, respectively. The SBHs of the InSe devices are estimated to be 70 meV, 190 meV, 280 meV, and 570 meV for In, Ti, Cr, and Pd contacts, respectively.

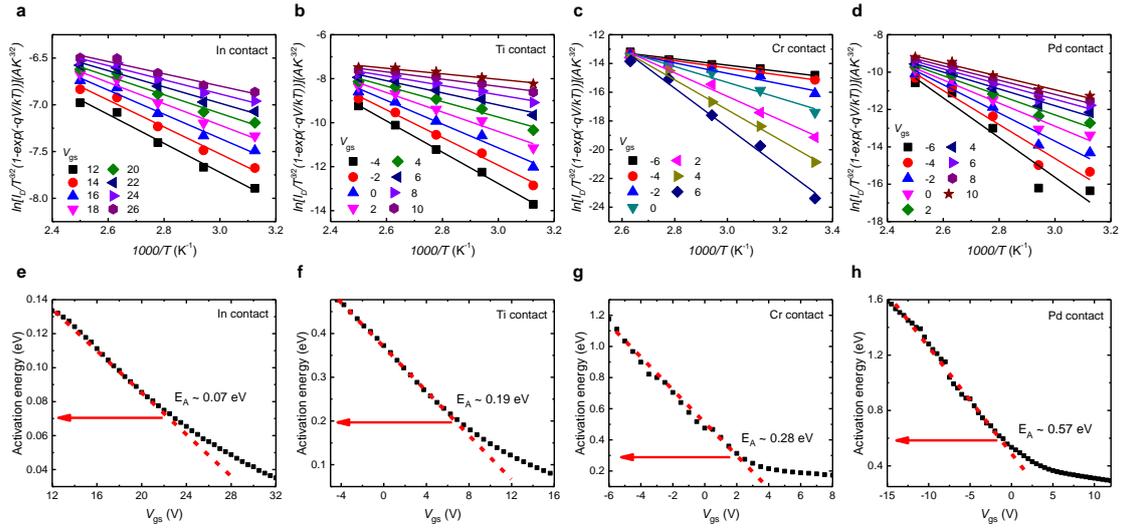

**Supplementary Figure S8.** Arrhenius plots of $I_{ds}/T^{3/2}$ and $E_A$ as functions of $V_{gs}$ for the OML-embedded InSe devices with (a, e) In, (b, f) Ti, (c, g) Cr, and (d, h) Pd contacts.

## S7. The OML-embedded InSe devices with an additional UV-ozone treatment

We show that the threshold voltage shifting observed in the OML-embedded InSe devices, as shown in Figure 4c in the main text, is irrelevant to the doping effect of the oxide layer. Supplementary Figure S10a shows a schematic of an OML-embedded InSe device with an additional UV-ozone treatment. For the additional UV-ozone treatment, the InSe device was oxidized under the same conditions at 80 °C for 10 seconds, leading to an additional OL in the channel region.

To confirm the origin of the change in the electrical behaviors, we performed a TEM characterization to clarify whether the oxidation process of InSe treated by UV ozone is self-limited. Supplementary Figure S9a shows a cross-sectional TEM image of an InSe sample treated with UV ozone at 80 °C for 20 sec. The UV-ozone-treated InSe with 20 sec grows to 2-nm-thick amorphous surface OL, about twice in thickness of the 10 sec process (Supplementary Figure S9b, same image as Figure 1a in the main text). This result clearly indicates that the oxidation by UV ozone is not a self-limiting process for InSe, at least in the first 20 sec. We can then infer that the second ozone exposure for the control InSe sample results in additional oxidized layer.

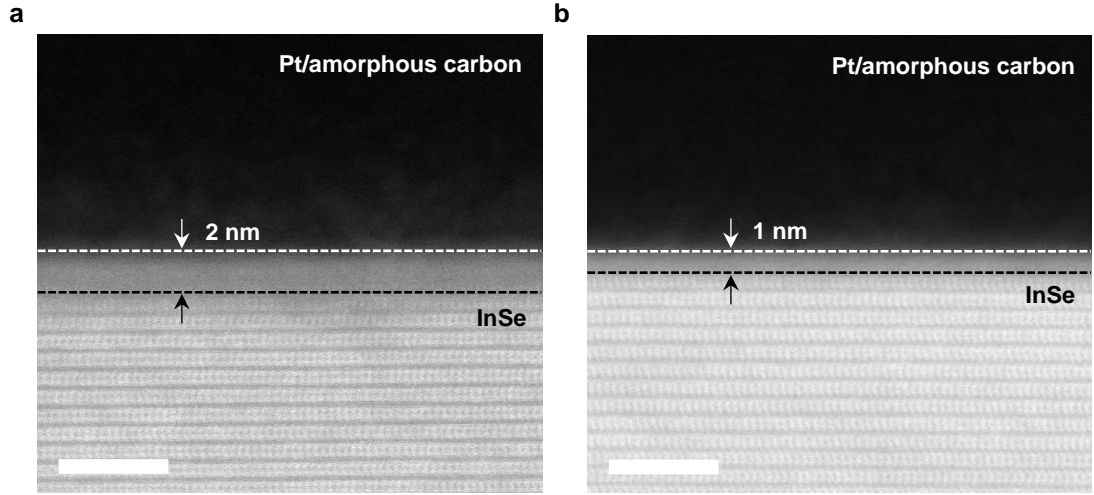

**Supplementary Figure S9.** Cross-sectional TEM images of InSe treated with UV ozone at 80 °C for 20 sec (Left) and 10 sec (Right). The scale bar is 5 nm.

Supplementary Figure S10b compares the transfer characteristics of the InSe devices before and after the UV-ozone treatment. This comparison shows negligible shift of $V_{th}$ in a control InSe device before and after an additional UV-ozone treatment, corresponding to 1 nm and 2 nm surface oxide layers, respectively. This TEM characterization supports that the shift in the $V_{th}$ shown in Figure 4b in the main text is improbable to be related to the doping effect of the oxide layer on the surface of 2D materials [8]. Moreover, as discussed in Figure 4b in the main text, $V_{th}$ shifts positively by 20 V and negatively by 25 V for larger (Pd) and smaller (In) WF metals, respectively. If the doping effect of the oxide layer were dominant, the shift of $V_{th}$ should manifest comparably. Based on the above discussion, we infer that the shift in $V_{th}$ is caused by the suppression of the Fermi level pinning as discussed in Figure 4b, and not due to the doping effect of the oxide layer.

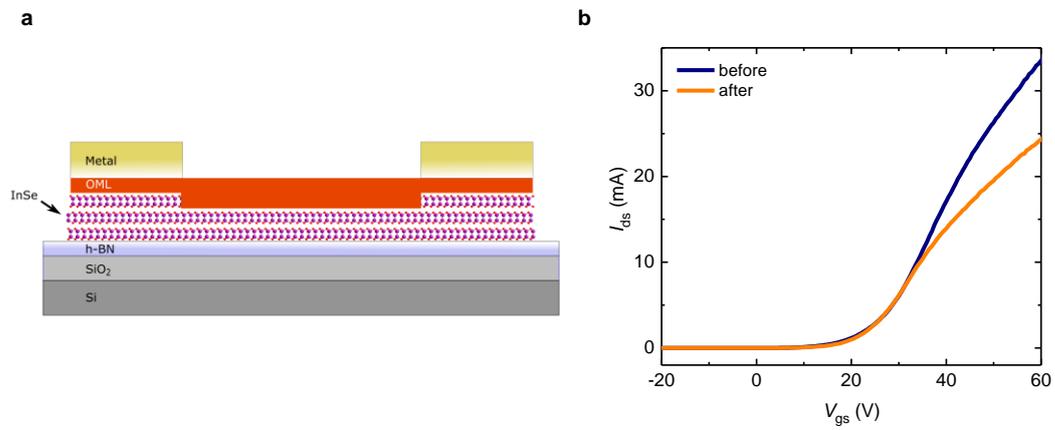

**Supplementary Figure S10.** (a) A schematic of an OML-embedded InSe device with an additional UV-ozone treatment. (b) Transfer characteristics of an InSe device before (black) and after (blue) the UV-ozone treatment. The $V_{ds} = 1$ V.

## S8. The distribution of the OML-embedded InSe devices and the control samples as a function of room-$T$ $\mu_{FE}$ and $N_{IT}$

To understand the dependence of the device performance on the surface trap density ($N_{IT}$), we plot the room-$T$ $\mu_{FE}$ distribution as a function of $N_{IT}$ for the OML-embedded InSe devices and the control samples without the UV-ozone treatment, as shown in Supplementary Figure S11. The $N_{IT}$ is calculated by extracting the difference in $V_t$ of the hysteresis,[11]

$$N_{IT} = C_{ox}\frac{\Delta V_t}{q},$$

where $C_{ox}$ is the gate capacitance and $\Delta V_t$ is the difference in the threshold voltage between the forward and the backward $V_{gs}$ sweeps. The OML-embedded InSe devices exhibit a higher room-$T$ $\mu_{FE}$ and smaller $N_{IT}$ ranging from $3 \times 10^{11}$ to $7 \times 10^{11}$ cm$^{-2}$. In contrast, the control samples show a lower room-$T$ $\mu_{FE}$ with a wide range of $N_{IT}$ ($3 \times 10^{11} - 2.5 \times 10^{12}$ cm$^{-2}$). The low mobility of the control samples can be attributed to the presence of charge scattering centers and trap states that limit the device performance.[12] Moreover, the larger $N_{IT}$ of the control samples may be ascribed to the native oxides, as evidenced by the AFM characterization in Supplementary Section 2. This sample statistics clearly indicate the enhanced device performance and the improved interface condition due to the integration of the OML.

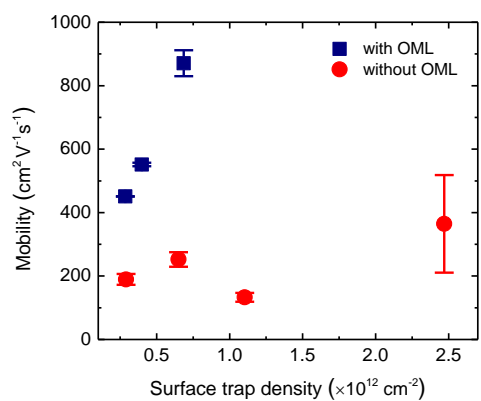

**Supplementary Figure S11.** The distribution of the OML-embedded InSe devices and the control samples as a function of $\mu_{FE}$ at room $T$ and $N_{IT}$. The error bar corresponds to the variation in $\mu_{FE}$ calculated from the forward and the backward $V_{gs}$ sweeps.

References

1   Ho, P.-H. *et al.* High-mobility InSe transistors: the role of surface oxides.   **11**, 7362-7370 (2017).

2   Balakrishnan, N. *et al.* Engineering p-n junctions and bandgap tuning of InSe nanolayers by controlled oxidation. *2d Mater* **4**, doi:Artn 025043
10.1088/2053-1583/Aa61e0 (2017).

3   Yamamoto, M. *et al.* Self-Limiting Layer-by-Layer Oxidation of Atomically Thin WSe2. *Nano Lett.* **15**, 2067-2073, doi:10.1021/nl5049753 (2015).

4   Yeh, Y.-C. *et al.* Growth of the Bi2Se3 Surface Oxide for Metal–Semiconductor–Metal Device Applications.   **120**, 3314-3318 (2016).

5   Yashina, L. V. *et al.* Negligible surface reactivity of topological insulators Bi2Se3 and Bi2Te3 towards oxygen and water.   **7**, 5181-5191 (2013).

6   Dean, C. R. *et al.* Boron nitride substrates for high-quality graphene electronics.   **5**, 722 (2010).

7   Zhao, Q., Frisenda, R., Wang, T. & Castellanos-Gomez, A. J. N. InSe: a two-dimensional semiconductor with superior flexibility.   (2019).

8   Simmons, J. G. J. J. o. a. p. Generalized formula for the electric tunnel effect between similar electrodes separated by a thin insulating film.   **34**, 1793-1803 (1963).

9   Simmons, J. G. J. J. o. A. P. Low-Voltage Current-Voltage Relationship of Tunnel Junctions.   **34**, 238-239 (1963).

10  Lee, S., Tang, A., Aloni, S. & Philip Wong, H.-S. J. N. l. Statistical study on the Schottky barrier reduction of tunneling contacts to CVD synthesized MoS2.   **16**, 276-281 (2015).

11  Yamamoto, M., Ueno, K. & Tsukagoshi, K. J. A. P. L. Pronounced photogating effect in atomically thin WSe2 with a self-limiting surface oxide layer.   **112**, 181902 (2018).

12  Han, C. *et al.* Oxygen induced strong mobility modulation in few-layer black phosphorus.   **4**, 021007 (2017).